\documentclass[review,12pt]{elsarticle}
\usepackage{lineno,hyperref}
\modulolinenumbers[5]

\journal{Journal of Templates}

\usepackage{amsmath}
\usepackage{graphicx}

\title{Thermoelectric Amplification of Phonons in Graphene}

\author[rvt]{K. A. Dompreh\corref{cor1}\fnref{fn1}}
\author[focal]{N. G. Mensah}
\author[rvt]{S. Y. Mensah}
\author[cvt  ]{S. K. Fosuhene}
\address[rvt]{Department of Physics, College of Agriculture and Natural Sciences, U.C.C, Ghana.}
\address[focal]{Department of Mathematics, College of Agriculture and Natural Sciences, U.C.C, Ghana}
\cortext[cor1]{Corresponding author\\ Email: kwadwo.dompreh@ucc.edu.gh}
\ead[url]{kwadwo.dompreh@ucc.edu.gh}
\address[cvt]{Ghana Space Science and Technology Institute. Ghana Atomic Energy Commission.}

\date{}
\begin{document} 
\begin{abstract}
Amplification of acoustic phonons  due to an external temperature gredient
($\nabla T$) in Graphene  was studied theoretically. The threshold temperature gradient 
$(\nabla T)_0^{g}$ at which absorption switches over to amplification in Graphene was 
evaluated at various frequencies ($\omega_q$)  and temperatures ($T$). For $T = 77K$  and 
frequency $\omega_q = 12THz$, $(\nabla T)_0^{g} = 0.37Km^{-1}$. The calculation was 
done in the regime at $ql >> 1$. The dependence of the normalized ($\Gamma/\Gamma_0$) on 
the frequency $\omega_q$ and the temperature 
gradient $(\nabla T/T)$ are evaluated numerically and presented graphically.  
The calculated $(\nabla T)_0^{g}$ for Graphene  is lower than that obtained for  
homogenous semiconductors ($n-InSb$) $(\nabla T)_0^{hom} \approx 10^3Kcm^{-1}$, 
superlattices $(\nabla T)_0^{SL} = 384Kcm^{-1}$, and cylindrical quantum wire $(\nabla T)_0^{cqw} \approx 10^2Kcm^{-1}$.
This makes Graphene a much better material for thermoelectric phonon amplifier.
\end{abstract}
\maketitle

\section*{Introduction}
The amplification (absorption) of acoustic phonons in Graphene~\cite{1,2,3} and other low dimensional
materials such as superlattices~\cite{4,5,6,7}, carbon nanotubes (CNT)~\cite{8} and cylindrical
quantum wires (CQW)~\cite{9} has attracted lots of 
attention recently. For Graphene, Nunes and Fonseca ~\cite{3} studied amplification 
of acoustic phonons and determined the drift 
velocity $V_D $ at which amplification occurs  but Dompreh et. al.~\cite{11} 
further showed that even at $V_D = 0$,
absorption of acoustic phonons can  occur. 
Acoustoelectric Effect (AE) involves the transfer of momentum from phonons to conducting charge 
carriers which leads to the generation of d.c. current in the sample.
This has been studied both theoretically~\cite{10,11} and experimentally~\cite{12} in Graphene.
The interaction between electrons and phonons in the presence of an external temperature gradient ($\nabla T$) can lead to 
thermoelectric effect~\cite{13,14,15,16} and thermoelectric amplification of phonons.
Thermoelectric amplification of phonons has been studied in bulk~\cite{17,18} and 
low dimensional materials  such as cylindrical quantum 
wire (CQW)~\cite{19} and superlattices~\cite{20,21}. 
This phenomena  was predicted  by Gulyeav and Ephstein (1967)~\cite{17} but was  thoroughly developed by 
Sharma and Singh (1974)~\cite{22} from a hydrodynamic approach $ql << 1$ ($q$ is the acoustic wave number, 
$l$ is the electron mean free path). Ephstein further explained this effect for 
sound in the opposite limiting case, $ql >> 1$ and   showed that  amplification is also
possible in an electrically open-circuited sample (i.e., in the absence of an electric current)~\cite{23}. 
In $n-InSb$, Epstein calculated a threshold gradient of $\approx 10^3Kcm^{-1}$ 
for a chemical potential $\mu = 8\times10^5cm^2V^{-1}s^{-2}$ at $77K$. However,   
in superlattices,  Mensah and Kangah ($1991$)~\cite{24}  calculated
the threshold temperature gradient necessary for amplification to occur  to be
$(\nabla T)_0^{hom} = 2.6(\nabla T)_0^{SL}$ (where $(\nabla T)_0^{SL}$ is the threshold temperature gradient 
of superlattice) but in the lowest energy miniband ($\Delta = 0.1eV$) obtained  $(\nabla T)_0^{SL} = 384 Kcm^{-1}$. 
In all these materials, the threshold temperature gradient  for the amplification was found to depend 
on the scattering mechanism and  sound frequency where the relaxation time is independent of 
energy~\cite{20}.\\
Graphene  differs significantly from the other low-dimensional materials. It has the highest value for thermal conductivity at 
room temperature ($\approx 3000-5000 W/{mK}$)~\cite{16,21}. This extremely high thermal conductivity opens
up a variety of applications. The most interesting property of Graphene is its  linear 
energy dispersion $E =\pm \hbar V_F \vert k \vert$ (the Fermi velocity $V_F \approx 10^8cms^{-1}$) at the Fermi level with 
low-energy excitation. At low temperatures, the conductivity in Graphene is restricted by 
scattering of impurities but in the absence of extrinsic scattering sources, phonons 
constitute an intrinsic source of scattering of electrons to produce measureable temperature difference 
($\nabla T$). Acoustic phonon scattering induced by low energy phonons  gives quantitatively 
small contribution of ($\nabla T$) in Graphene even at room temperature. This is due to the high Fermi temperature of 
Graphene~\cite{25}. To-date, there is no study of thermoelectric amplification of acoustic phonons in Graphene.
In this paper, the effect is theoretically studied  in 
Graphene  with degenerate energy dispersion. Here the threshold temperature 
gradient  $(\nabla T)_0^{g}$ above which amplification occur is calculated in the regime $ql >> 1$.
Furthermore the frequency at which the graphs converges are calculated. 
The paper is organised as follows: In the theory section, 
the equation underlying the thermoelectric amplification of acoustic phonon in Graphene is 
presented. In the  numerical analysis section,  the final equation is analysed  and 
presented in a  graphical form.   Lastly, the  conclusion is presented in section $4$.    
\section*{Theory}
The kinetic equation for the acoustic 
phonon population $N_{\vec{q}}(t)$ in the Graphene sheet is given by
\begin{eqnarray}
\frac{\partial N_{\vec{q}}(t)}{\partial t} =\frac{2\pi}{\hbar}g_s g_v\sum_{k,k^{\prime}}\vert{C_{\vec{q}}}\vert^2 \delta_{k,k^{\prime}+{\vec{q}}}
\{[N_{\vec{q}}(t) + 1]f_{\vec{k}}(1-f_{\vec{k}^\prime})
\delta(\varepsilon_{\vec{k}^\prime} - \varepsilon_{\vec{k}} +\hbar\omega_{\vec{q}})\nonumber\\
- N_{\vec{q}}(t) f_{\vec{k}^\prime}(1-f_{\vec{k}})\delta(\varepsilon_{\vec{k}^\prime} - \varepsilon_{\vec{k}} - \hbar\omega_{\vec{q}})\}
\end{eqnarray}
where $g_s = g_v = 2$ account the for spin and valley degeneracy respectively, $N_{\vec{q}}(t)$ represent the number of phonons with a wave vector $\vec{q}$ at time $t$. 
The factor $N_{\vec{q}} + 1$ accounts for the 
presence of $N_{\vec{q}}$ phonons in the system when the additional phonon is emitted. The $f_{\vec{k}}(1-f_{\vec{k}})$ represent the probability that 
the initial $\vec{k}$ state is occupied and the final electron state $\vec{k}^\prime$ is empty whilst the factor $ N_{\vec{q}} f_{\vec{k}^\prime}(1-f_{\vec{k}})$
is that of the boson and fermions statistics.  In Eqn ($1$), the summation over $k$ and $k^{\prime}$ can be transformed into integrals by the prescription
$$\sum_{k,k^\prime}\rightarrow \frac{A^2}{(2\pi)^4}\int d^2 k d^2k^{\prime}$$
where $A$ is the area of the sample, and assuming that $N_q(t) >> 1$ yields
\begin{equation}
\frac{\partial N_{\vec{q}}}{\partial t} = \Gamma_{\vec{q}}N_{\vec{q}}
\end{equation}
where 
\begin{eqnarray}
\Gamma_q = \frac{ A\vert \Lambda\vert^2}{(2\pi)^3\hbar V_F\rho V_s}\int_0^\infty{kdk}\int_0^{\infty}{k^\prime dk^\prime}
\int_0^{2\pi}{d\phi}\int_0^{2\pi}{d\theta}\{[f(k)-f(k^\prime)]\times\nonumber \\
\delta(k-k^\prime-\frac{1}{\hbar V_F}(\hbar\omega_q ))\} 
\end{eqnarray}
with  $ k^\prime = k - \frac{1}{\hbar V_F}(\hbar\omega_q )$. 
${\Lambda}$ is the constant of deformation potential,
$\rho$ is the density of the Graphene sheet. $f(k)$ is the distribution function, $V_s$ is the velocity of sound, 
and $A$ is the area of the Graphene sheet. Here the acoustic wave will be considered as phonons of frequency ($\omega_q$) in 
the short-wave region  $ql >> 1$ ($q$  is the acoustic wave number, $l$ is the electron mean free path). 
From Eqn.($3$), the linear approximation of the distribution function $f(k)$ is given as 
\begin{equation}
f(k) = f_0(\varepsilon(k)) + q f_1(\varepsilon(k))
\end{equation}
At low temperature $k_B T << 1$, the Fermi-Dirac equilibrium distribution function become 
\begin{equation}
f_0(\varepsilon(k)) = exp(-\beta(\varepsilon(k))) 
\end{equation}
From Eqn. (4), $f_1(k)$ is derived from the Boltzmann transport 
equation as 
\begin{equation}
f_1(\varepsilon(k)) = \tau[(\varepsilon(k) - \xi)\frac{\nabla T}{T}]\frac{\partial f_0(p)}{\partial\varepsilon}v(k)
\end{equation}
Here $v(k)= {\delta\varepsilon(k)}/{\hbar\delta k}$ is the electron velocity, $\xi$ is the chemical potential 
$\tau$ is the relaxation time and $\nabla T$ is the temperature difference.  
Inserting Eqn.($4$), Eqn.($5$) and Eqn.($6$) into Eqn.($3$) and expressing further gives
\begin{eqnarray}
\Gamma=\frac{e A\vert \Lambda\vert^2}{(2\pi)V_F\rho V_s}\int_0^\infty(k^2 -\frac{k\hbar\omega_q}{\hbar V_F})
\{exp(-\beta(\hbar V_F k ))-\beta\hbar V_F q\tau(\hbar V_F k)\times\nonumber\\
 \frac{\nabla T}{\hbar T}exp(-\beta\hbar V_F k)- exp(-\beta \hbar V_F (k - \frac{\hbar\omega_q}{\hbar V_F})) -
\beta \hbar {V_F} q \tau(\hbar V_F(k -\frac{\hbar\omega_q}{\hbar V_F}))\times\nonumber\\
\frac{\nabla T}{\hbar T}exp(-\beta \hbar V_F(k -\frac{\hbar\omega_q}{\hbar V_F}))\}
\end{eqnarray}
Using standard integrals in Eqn($7$) and after a cumbersome calculation
yields the final  equation  as 
\begin{eqnarray}
\Gamma &=& \Gamma_0 \{(2 -\beta\hbar\omega_q)(1- exp(-\beta\hbar\omega_q))\nonumber\\
&-& q\tau V_F [6(1+exp(\beta\hbar\omega_q)) -\beta\hbar\omega_q(2 + \beta\hbar\omega_qexp(\beta\hbar\omega_q))]\frac{\nabla T}{T}\}
\end{eqnarray}
where 
\begin{equation}
\Gamma_0 =  \frac{2A\vert {\Lambda}\vert^2 q}{2\pi{\beta}^3{\hbar}^3{ V_F}^4\rho V_s}\label{Eq_6}
\end{equation}
The threshold temperature difference $(\nabla T)_0^{g}$ is calculated. This is 
achieved by letting $\Gamma = 0$ as a consequence of the laws of conservation which yields 
\begin{equation}
(\nabla T)_0^g = 
T\frac{\{(2 -\beta\hbar\omega_q)(1- exp(-\beta\hbar\omega_q))\}}
{q\tau V_F [6(1+exp(\beta\hbar\omega_q)) -\beta\hbar\omega_q(2 + \beta\hbar\omega_qexp(\beta\hbar\omega_q))]}
\end{equation}

\section*{Numerical Analysis}
To understand the complex expression for the absorption, Eqn($8$) was numerically analysed.
From Eqn. ($10$), the threshold temperature gradient$(\nabla T)_0^g$ is dependent on the temperature ($T$), the frequency 
($\omega_q$) and the relaxation time ($\tau$) as well as the acoustic wavenumber ($q$). To evaluate $(\nabla T)_0^g$, 
the following parameters are used $\Lambda = 9 eV$, $V_s = 2.1\times 10^3ms^{-1}$, $\tau =10^{-12}s$, 
$\omega_q = 10^{12} s^{-1}$ and ${q} = 10^8 m^{-1}$. At $T = 77 K$, $(\nabla T)_0^{g}$ is calculated to be $0.37 Km^{-1}$. 
These values are used in analysing Eqn.($8$) numerically and presented graphically (see Figure $1$, $2$ and $3$). In Figure $1a$,  
A normalized graph of $\Gamma/\Gamma_0$  on  $\omega_q$ for varying $\nabla T$ is plotted. When $\nabla T = 0$, an absorption 
graph ($\Gamma/\Gamma_0 > 0$) is obtained. We observed that when $\nabla T = 0.28 Km^{-1}$, the amplitude of the absorption
graph reduces and the absorption  switches to amplification at $\omega_q = 10THz$.
Interestingly, when $\nabla T  > (\nabla T)_0^{g} = 0.37Km^{-1}$, the graph switches 
completely to  amplification  ($\Gamma/\Gamma_0 < 0$) 
(see Figure $1a$ for graph of $\nabla T = 0.38 Km^{-1}$ and $\nabla T = 0.48 Km^{-1}$). This agrees with the theory of 
thermoelectric amplification of phonons. 
\begin{figure}[ht!]
\begin{centering}
\includegraphics[width = 9cm]{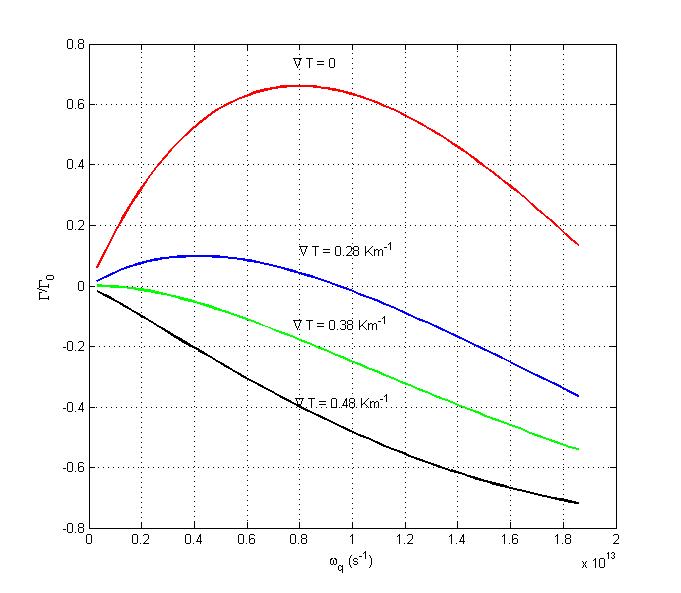}
\caption{(a) The normalized graph of  $\Gamma/\Gamma_0$ on $\omega_q$  for varying $\nabla T$} 
\end{centering}
\end{figure}
\begin{figure}[ht!]
\includegraphics[width =7.5cm]{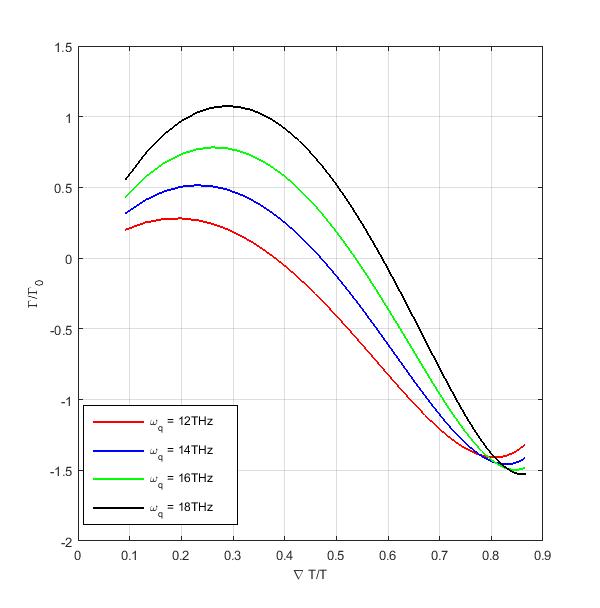}
\includegraphics[width =7.5cm]{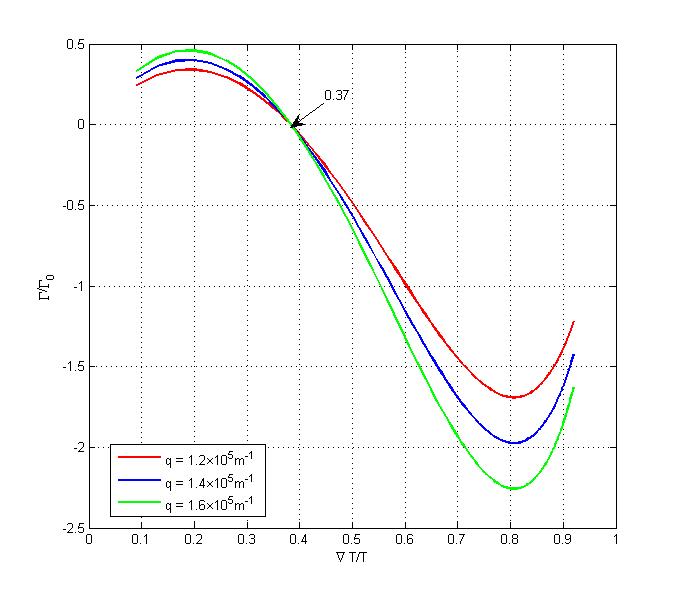}
\caption{The normalized graph of  $\Gamma/\Gamma_0$ on $\nabla T/T$ (a) for varying $\omega_q$  
(b) for varying $q$} 
\end{figure}
\begin{figure}[ht!]
\includegraphics[width = 7.5cm]{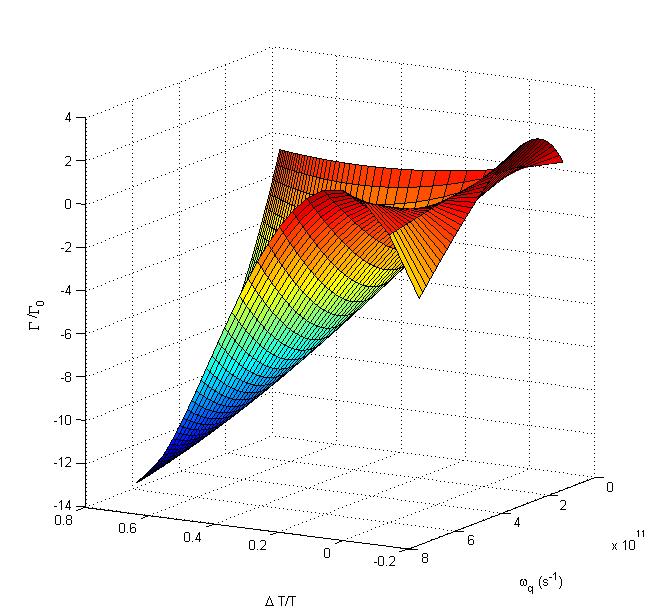}
\includegraphics[width = 7.5cm]{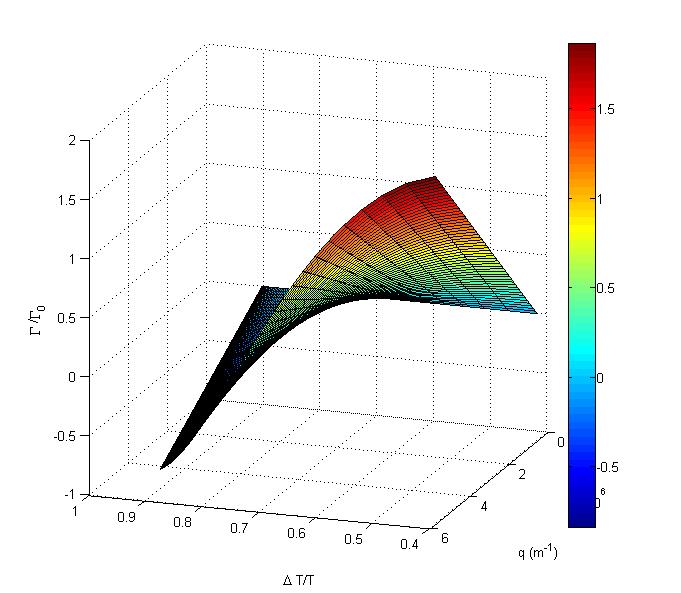}
\caption{ The Dependence of $\Gamma/\Gamma_0$ on $\omega_q$  and $\nabla T/T$.}
\end{figure}
\begin{figure}[ht!]
\begin{centering}
\includegraphics[width =10cm]{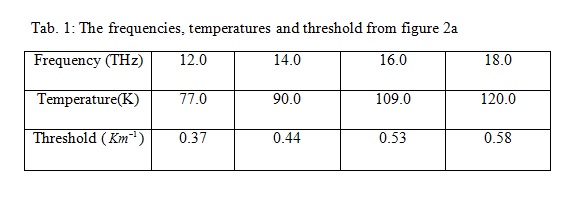}
\end{centering}
\end{figure}
The threshold values above which absorption completely switches to amplification are 
presented in table $1$ for various temperatures and frequencies. In the table $1$, 
$(\nabla T)_0^{g}$ increases with increase in temperature and frequency.  
 Figure $1b$ shows the 
graph of various $\Gamma/\Gamma_0$ on $\nabla T/T$ for varying $q$. By increasing  $q$, all the graphs switches to amplification 
at the same point ($\nabla T/T = 0.37 $). This shows that, increases in the acoustic wavenumber ($q$) does not alter 
the threshold temperature gradient$(\nabla T)_0^{g}$ of the material. However, when
the graph of  $\Gamma/\Gamma_0$ was plotted against $\nabla T/T$ varying the frequency, we notice a shift of the 
threshold values towards bigger values of $\nabla T/T$ (see Figure $2b$).
A $3D$ graph is presented to further elucidate the graphs obtained (see Figure $3$). 
Figure ($3a$) shows the dependence of $\Gamma/\Gamma_0$ on $\nabla T/T$ and  $\omega_q$
whilst Figure ($3b$) is that of $\Gamma/\Gamma_0$ on $\nabla T/T$ and  $q$.
  
\section*{Conclusion}
Thermoelectric amplification of phonons in Graphenes is studied. We observed that absorption
switches over to amplification at  values greater than the threshold values.
The  threshold value calculated  at $T = 77K$  and frequency $\omega_q = 10^{12}s^{-1}$ in Graphene is 
$(\nabla T)_0^{g} = 0.37 Kcm^{-1}$ which is far lower than that calculated in  homogenous 
semiconductor using $n-InSb$ ($s = 2.3\times10^5cms^{-1}$, $\mu = 8\times10^5 cm^2V^{-1}s^{-1}$
at $77K$) and was found to be $(\nabla T)_0^{hom}\approx 10^3Kcm^{-1}$~\cite{17},  
superlattice $(\nabla T)_0^{SL} \approx 384Kcm^{-1} $ for miniband width $\Delta = 0.1eV$ at
$77K$~\cite{20}, and finally for 
cylindrical quantum wires (CQW) to be $(\nabla T)_0^{cwq}\approx 10^2Kcm^{-1}$  at liquid 
Nitrogen temperature of $77K$~\cite{19}. This makes Graphene a much better material for 
thermoelectric  phonon amplification.

\renewcommand\refname{Bibliography}


\begin{thebibliography}{99} 
\bibitem{1}
Dompreh, Kwadwo A., Natalia G. Mensah, and Samuel Y. Mensah. "Amplification 
of Hypersound in Graphene with degenerate energy dispersion." arXiv preprint 
arXiv:1503.07360 (2015).
\bibitem{2}
Dompreh, K. A., S. Y. Mensah, S. S. Abukari, F. Sam, and N. G. Mensah. "Amplification of 
Acoustic Waves in Graphene Nanoribbon in the Presence of External Electric and Magnetic 
Field." arXiv preprint arXiv:1410.8064 (2014).
\bibitem{3}
Nunes, O. A. C., and A. L. A. Fonseca. "Amplification of hippersound in graphene under 
external direct current electric field." Journal of Applied Physics 112, no. 4 (2012): 043707.
\bibitem{4}
Shmelev, G. M., S. Y. Mensah, and G. I. Tsurkan. "Hypersound amplification by a superlattice in 
a nonquantised electric field." Journal of Physics C: Solid State Physics 21, no. 33 (1988): 
L1073.
\bibitem{5}
Mensah, S. Y., and G. K. Kangah. "Amplification of acoustic waves due to an external 
temperature gradient in superlattices." Journal of Physics: Condensed Matter 3, no. 22 
(1991): 4105.
\bibitem{6}
Mensah, S. Y., F. K. A. Allotey, N. G. Mensah, and V. W. Elloh. "Amplification of acoustic 
phonons in a degenerate semiconductor superlattice." Physica E: Low-dimensional Systems and 
Nanostructures 19, no. 3 (2003): 257-262.
\bibitem{7}
Dompreh, Kwadwo A., Samuel Y. Mensah, Natalia G. Mensah, Sulley S. Abukari, Frank KA Allotey, 
and George K. Nkrumah-Buandoh. "Amplification of acoustic phonons in superlattice." arXiv 
preprint arXiv:1101.1854 (2011).
\bibitem{8}
Dompreh, K. A., N. G. Mensah, S. Y. Mensah, S. S. Abukari, F. Sam, and R. Edziah. "Hypersound 
Absorption of Acoustic Phonons in a degenerate Carbon Nanotube." arXiv preprint arXiv:1502.07636 (2015).
\bibitem{9}
Hung, Nguyen Quoc, Nguyen Vu Nhan, and Nguyen Quang Bau. "On the amplification of sound 
(acoustic phonons) by absorption of laser radiation in cylindricalquantum wires." arXiv 
preprint cond-mat/0204563 (2002).
\bibitem{10}
Zhao, C. X., W. Xu, and F. M. Peeters. "Cerenkov emission of terahertz acoustic-phonons from 
graphene." Applied Physics Letters 102, no. 22 (2013): 222101.
\bibitem{11}
Dompreh, K. A., N. G. Mensah, and S. Y. Mensah. "Acoustoelectric Effect in Graphene with 
degenerate Energy dispersion." arXiv preprint arXiv:1505.05031 (2015).
\bibitem{12}
Bandhu, L., L. M. Lawton, and G. R. Nash. "Macroscopic acoustoelectric charge transport in graphene." 
Applied Physics Letters 103, no. 13 (2013): 133101.
\bibitem{13}
Mensah, S. Y., F. K. A. Allotey, N. G. Mensah, and G. Nkrumah. "Differential thermopower of a CNT chiral 
carbon nanotube." Journal of Physics: Condensed Matter 13, no. 24 (2001): 5653.
\bibitem{14}
Mensah, S. Y., F. K. A. Allotey, N. G. Mensah, and G. Nkrumah. "Giant electrical power factor in single-walled 
chiral carbon nanotube." superlattices and microstructures 33, no. 3 (2003): 173-180.
\bibitem{15}
Mensah, S. Y., A. Twum, N. G. Mensah, K. A. Dompreh, S. S. Abukari, and G. Nkrumah-Buandoh. "Effect of laser on 
thermopower of chiral carbon nanotube." arXiv preprint arXiv:1104.1913 (2011).
\bibitem{16}
Mensah, N. G., G. Nkrumah, S. Y. Mensah, and F. K. A. Allotey. "Temperature dependence of the thermal conductivity 
in chiral carbon nanotubes." Physics Letters A 329, no. 4 (2004): 369-378.
\bibitem{17}
Gulayev, Yu V. "Acousto-electric effect and amplification of sound waves in 
semiconductors at large sound intensities." Physics Letters A 30.4 (1969): 260-261.
\bibitem{18}
Tenan, M. A., A. Marotta, and L. C. M. Miranda. "On the thermoelectric amplification of sound in semiconductors.
" Applied Physics Letters 35, no. 4 (1979): 321-323.
\bibitem{19}
Nunes, O. A. C., D. A. Agrello, and A. L. A. Fonseca. "Thermoelectric amplification of phonons in cylindrical
quantum well wires." Journal of applied physics 83 (1998): 87-89.
\bibitem{20}
Mensah, S. Y., and G. K. Kangah. "The thermoelectric effect in a semiconductor superlattice in a non-quantized 
electric field." Journal of Physics: Condensed Matter 4, no. 3 (1992): 919.
\bibitem{21}
Balandin AA. Thermal properties of graphene and nanostructured carbon materials.
Nat Mat 2011;10(8):569–81
\bibitem{22}
Sharma, S. and Singh, S. P., J. Appl Phys. 10 46656-1 (1974)
\bibitem{23}
Epstein E.M. Fiz. Tekh. Poluprov. 8 1584-7 (1975)
\bibitem{24}
Mensah, S. Y., and G. K. Kangah. "Amplification of acoustic waves due to an external temperature gradient in superlattices." 
Journal of Physics: Condensed Matter 3, no. 22 (1991): 4105.
\bibitem{25}
Sankeshwar, N. S., S. S. Kubakaddi, and B. G. Mulimani. "Thermoelectric power in graphene." Intech 9 (2013): 1-56.

\end{thebibliography}
\end{document}